# Transport in Nanoribbon Interconnects Obtained from Graphene Grown by Chemical Vapor Deposition

Ashkan Behnam[*1,2], Austin S. Lyons[*1,2], Myung-Ho Bae[1,2,3], Edmond K. Chow[1], Sharnali Islam[1,2], Christopher M. Neumann[1,2], and Eric Pop[1,2,4]

[1]*Micro and Nanotechnology Lab, Univ. Illinois at Urbana-Champaign, Urbana, IL 61801, USA*
[2]*Dept. of Electrical & Computer Eng., Univ. Illinois at Urbana-Champaign, Urbana, IL 61801, USA*
[3]*Present address: Division of Convergence Technology, Korea Research Institute of Standards and Science, Daejeon 305-340, Republic of Korea*
[4]*Beckman Institute for Advanced Studies, Univ. Illinois at Urbana-Champaign, Urbana, IL 61801, USA*

[*] These authors contributed equally to this work

Contact: epop@illinois.edu

**ABSTRACT:** We study graphene nanoribbon (GNR) interconnects obtained from graphene grown by chemical vapor deposition (CVD). We report low- and high-field electrical measurements over a wide temperature range, from 1.7 to 900 K. Room temperature mobilities range from 100 to 500 $cm^2V^{-1}s^{-1}$, comparable to GNRs from exfoliated graphene, suggesting that bulk defects or grain boundaries play little role in devices smaller than the CVD graphene crystallite size. At high-field, peak current densities are limited by Joule heating, but a small amount of thermal engineering allows us to reach ~$2 \times 10^9$ $A/cm^2$, the highest reported for nanoscale CVD graphene interconnects. At temperatures below ~5 K, short GNRs act as quantum dots with dimensions comparable to their lengths, highlighting the role of metal contacts in limiting transport. Our study illustrates opportunities for CVD-grown GNRs, while revealing variability and contacts as remaining future challenges.

Graphene nanoribbons (GNRs) are promising candidates for nanoelectronics building blocks as interconnects, transistors, or sensors[1-4]. Previous studies have characterized individual GNRs prepared from chemically derived[1, 2, 5], mechanically exfoliated[4, 6], or epitaxially grown[7] graphene. However, these fabrication methods are less practical or more expensive for future large-scale integrated circuit fabrication. On the other hand, chemical vapor deposition (CVD) has been used as a facile approach for synthesizing large area polycrystalline graphene films[8-10] with grain sizes from tens of nanometers to microns[11, 12]. CVD-grown graphene has been recently investigated as a promising material for micron-sized interconnects, either on CMOS[13, 14] or on transparent and flexible substrates[15]. However, nanometer scale GNR interconnects from CVD graphene have not been systematically studied to date. Such GNRs represent the ultimate scaling limits of graphene interconnects and could be comparable to or smaller than the average CVD graphene crystallite size, leading to few or no bulk defects in individual devices. This could achieve the dual purpose of large-scale fabrication with relatively good quality GNRs.

In this work we present a comprehensive analysis of nanoscale GNRs with widths $W < 100$ nm and lengths $L < 800$ nm obtained from patterned CVD graphene. We find that such CVD GNRs have electrical properties comparable to those obtained by any other methods, suggesting a negligible effect of bulk defects or grain boundaries on their performance. At high fields we attain some of the highest current densities recorded in either graphene or GNR interconnects (~2 $\times 10^9$ A/cm$^2$); at low temperatures we note evidence of single quantum dots with size comparable to the channel length, underlining how contacts determine the conductance levels in such nanoscale devices. This study also serves to identify future challenges, and represents a fundamental stepping stone towards large-scale integration of nanoscale GNR interconnects.

Our CVD graphene growth and GNR device process steps are illustrated in Figure 1 and in the Supplementary Information. Briefly, graphene is grown on Cu foil[8] then transferred to SiO$_2$ (90 nm) on Si substrates (n+ doped), and annealed to remove water and organic residue. We define large Ti/Au (0.5/40 nm) contact pads (Figure S1 in Supplement) using optical lithography, followed by smaller finger electrodes by electron-beam (e-beam) lithography. We then define and deposit a narrow "strip" of Al (2-4 nm thick) which serves as the etch mask for the GNRs (Figure 1c and Figure S1b). The thin Al oxidizes when the chip is removed from the evaporation chamber and lift-off leaves behind an AlO$_x$ nanoribbon covering the graphene and stretching between the finger electrodes (Figure S1c). This AlO$_x$ strip serves the multiple purpose of protect-ing the graphene, serving as a dielectric seeding layer, and being scalable for large-area fabrication. (Some, albeit not all these goals could also be achieved with a thicker ~20 nm metal strip[16] or a nanowire mask[17, 18] for etching the GNRs.) The GNRs are defined by a short $O_2$ plasma etch which removes the unprotected graphene. The graphene which fans out under the contacts (Figure 1b) is protected during the etch, helping manage contact resistance. The $AlO_x$ strip was left on some devices (batch b1), and removed on others (batch b2) for the measurements.

Figure 1f compares the Raman spectrum of the unpatterned CVD graphene to several individual GNRs. All GNRs display the disorder-induced Raman D peak and most also display the D' peak, which are accentuated in GNRs due to the presence of edges[19, 20]. The integrated D to G peak area ratio of our GNRs from CVD graphene is $A_D/A_G$ ~1-5, comparable to $A_D/A_G$ ~2-8 measured for arrays of GNRs from exfoliated graphene of similar widths[19, 20]. For comparison, our bulk CVD graphene has $A_D/A_G$ ~ 0.2. Using this ratio we can calculate the average crystallite size following Cancado et al[21], $L_a$ ~ 200 nm or an average area ~$4\times10^4$ $nm^2$. The length scale $L_a$ approximately corresponds to the average distance between defects (including bulk defects and grain boundaries), thus it is smaller than the polycrystalline grain size in the CVD-grown graphene. Similar values were also recently estimated by scanning tunneling microscopy (STM) in our group, on comparable CVD graphene growths[22]. Thus, given GNR areas from $0.2\times10^4$ to $4\times10^4$ $nm^2$ in this work, it is highly likely that most samples are monocrystalline and free of bulk defects.

Low-bias measurements of both $AlO_x$-capped and bare CVD GNRs in air reveal similar $p$-doping (Supplement Figure S2). Transferring devices to a vacuum probe station (~$10^{-5}$ Torr) and annealing at 300 ℃ for 2 hours removes most of the physisorbed ambient impurities such as water[23], oxygen[24] and PMMA residue[25]. After annealing, measurements in vacuum show devices are less $p$-doped than in air and in some cases $n$-doped (Figure S2b). We note that our vacuum probe station has high-temperature capability, enabling electrical measurements after sample anneal, without breaking vacuum. We fit all electrical data with a transport model[26, 27] which includes the gate dependence ($V_G$), thermally generated carriers ($n_{th}$), puddle charge ($n_{pd}$) due to substrate impurities, and contact resistance ($R_C$) effects. (More information is provided in Section C of the Supplement.) Since GNRs are narrow compared to the underlying oxide thickness, the effect of fringing fields on the capacitance must be included[2, 28] (Figure 2a inset). Thus, we use an expression for the capacitance per unit area as[2]:

$$C_{ox} \approx \varepsilon_{ox} \varepsilon_0 \left\{ \frac{\pi}{\ln\left[ 6\left( t_{ox}/W + 1 \right) \right] W} + \frac{1}{t_{ox}} \right\}, \tag{1}$$

where $t_{ox} \approx 90$ nm is the SiO$_2$ thickness, $\varepsilon_{ox} \approx 3.9$ is the relative permittivity of SiO$_2$ and $\varepsilon_0 \approx 8.854 \times 10^{-14}$ F/cm is the permittivity of vacuum. The first term represents the fringing capacitance and the second term is the parallel plate capacitance between the GNR and the top of the n+ Si substrate. As an example, for a GNR with $W = 40$ nm on $t_{ox} = 90$ nm, ~72% of the total capacitance is due to fringing fields and the rest due to parallel plate capacitance. In the limit $W \gg t_{ox}$, the equation reduces to the usual $C_{ox} = \varepsilon_{ox} \varepsilon_0 / t_{ox}$ as expected, and quantum capacitance[29] can be neglected due to the thickness of the buried oxide. (Figure S3 in the Supplement illustrates the contribution of fringing capacitance to the total capacitance as a function of $W$.)

Fitting our model against the experimental data reveals a mobility range $\mu \approx 100$-500 cm$^2$V$^{-1}$s$^{-1}$ and contact resistance $R_C W \geq 500$ Ω·μm at room temperature (per width), for these GNRs obtained from CVD-grown graphene.[30] The mobility values of our CVD GNRs are comparable to lithographically patterned GNRs from exfoliated graphene[4] (100~1000 cm$^2$V$^{-1}$s$^{-1}$) and somewhat lower than GNRs from unzipped nanotubes[1,2,31] (100~3200 cm$^2$V$^{-1}$s$^{-1}$), ostensibly due to lesser edge disorder of the latter. However, the similarity of mobility for exfoliated vs. CVD-grown graphene does not exist for larger samples, as micron-scale (polycrystalline) CVD graphene devices consistently show lower mobility than (crystalline) exfoliated ones[26,32]. This suggests that bulk defects or grain boundaries play almost no role in lowering our GNR mobility, a consequence of the nanoscale GNR dimensions being comparable to or smaller than the crystallite size of CVD-grown graphene. Nevertheless, the mobility values of our GNRs from CVD graphene remain lower than those of large samples (Figure 2a), illustrating that transport is still limited by edge roughness scattering, which must be better controlled in future work.

In order to further understand the transport properties of our GNRs, we undertook temperature-dependent measurements on several samples, as shown in Figure 2. GNR mobility or contact resistance data over a wide temperature range have not been available until now, to our knowledge. Figure 2a displays extracted mobility from three GNRs and two large-area devices (500×100 and 500×75 μm respectively) from a comparable CVD graphene growth[32]. The large devices show mobilities that are 3–6 times higher than those obtained for GNRs, and are likely limited by surface impurities and grain boundary or defect scattering[33], as the device size is much greater than the crystallite size $L_a$. On the other hand the GNRs are smaller than $L_a$, thus their lower mobility is attributed primarily to edge scattering, although differences in surface impurities between samples cannot be ruled out and could explain the variability noted.

The GNRs with lower mobility (~100 cm$^2$/V·s) in Figure 2a show virtually no temperature dependence. This is consistent with a transport regime where scattering rates from acoustic phonons, surface impurities and edge roughness are nearly equal and their opposite temperature dependence cancels out[34]. However, the ~20-nm-wide GNR with higher mobility shows a slight increase up to room temperature[30], consistent with a transport regime limited by scattering from surface impurities[34, 35]. The mobility then transitions to a weakly phonon-limited regime above room temperature, indicating that this device may be approaching the upper, intrinsic limits of achievable transport in GNRs of this width and edge roughness on SiO$_2$. Its mobility is also at the upper end of what was achieved in GNRs of this width patterned from exfoliated graphene[4].

The total resistance ($R$) and contact resistance ($R_C$) dependence on temperature for this sample are shown in Figure 2b and its inset. The $R_C$ dependence on temperature and carrier density is given through its dependence on sheet resistance $R_S$ (see Supplement Section C); thus, $R_C$ for such GNRs is almost independent of temperature like the mobility, but it scales approximately as the inverse square root of carrier density, $\propto (n + p)^{-1/2}$ (also see refs. 27, 36). The uncertainty in the $R_C$ extraction arises partly from the fitting algorithm (as for mobility) and partly from uncertainty of the GNR width which fans out under the metal contact, taken here between $W$ and $W + 2L_T$ where $L_T$ is the current transfer length into the contact electrode[36]. It is important to note that our model includes thermally generated carriers[26] ($n_{th}$), which are sometimes neglected but turn out to be crucial in fitting the correct temperature-dependent behavior of the graphene conductance at room temperature and above.

We now turn to the high-field behavior of our GNR interconnects, to understand their maximum current-carrying capacity up to electrical breakdown (BD). Figure 3 shows the results of high-field measurements at room temperature for 22 GNRs with widths $W = 15-50$ nm and lengths $L = 100-700$ nm. Figure 3a shows representative current-voltage data obtained from four GNRs in air. Figure 3b suggests that the GNR breakdown power ($P_{BD}$) scales approximately with the square root of the GNR area, a first indication of the role of heat dissipation from GNRs to the substrate[2, 37]. Recasting our measured data as breakdown current density ($J_{BD}$) vs. resistivity ($\rho$) in Figure 3c, we find scaling similar to both large-area CVD graphene interconnects[14] and GNRs from exfoliated graphene[38]. However, for a given resistivity, the current density of our GNRs from CVD graphene on 90 nm SiO$_2$ exceeds that of previously measured samples on 300 nm SiO$_2$[14, 38]. To understand these scaling relationships, we apply the model of Liao et al[2] which includes both heat loss to the substrate and to the contacts:

$$J_{BD} = \left[ \frac{g(T_{BD} - T_0)}{\rho \, t_g W} \times \frac{\cosh\left(\dfrac{L}{2L_H}\right) + gL_H R_T \sinh\left(\dfrac{L}{2L_H}\right)}{\cosh\left(\dfrac{L}{2L_H}\right) + gL_H R_T \sinh\left(\dfrac{L}{2L_H}\right) - 1} \right]^{1/2}.$$ (2)

Here, $T_{BD}$ is the breakdown temperature ($\sim$600 $^o$C oxidation in air), $T_0$ is the ambient temperature (22 $^o$C), $t_g$ is the thickness of the GNR, $L_H = (k_g W t_g / g)^{1/2}$ is the thermal healing length[2, 39] and $k_g$ the thermal conductivity along the GNR, and $R_T$ is the thermal resistance at the metal contacts[2]. The first term in eq. 2 above accounts for heat sinking into the substrate, and the second term accounts for heat sinking into the contacts; in the limit $L \gg L_H$ the second term becomes unity, i.e. heat sinking through the substrate dominates for long GNRs. (Complete model information is provided in Section F of the Supplement.) The thermal resistance per unit length from the GNR to the substrate is calculated as[2, 37]

$$g^{-1} = \left\{ \frac{\pi k_{ox}}{\ln\left[6(t_{ox}/W + 1)\right]} + \frac{k_{ox}}{t_{ox}} W \right\}^{-1} + \frac{R_{Cox}}{W} + \frac{1}{2k_{Si}}\left(\frac{L}{W_{eff}}\right)^{1/2},$$ (3)

where $k_{ox}$ = 1.4 Wm$^{-1}$K$^{-1}$ is the thermal conductivity of SiO$_2$, $R_{Cox}$ is the thermal resistance of the graphene-SiO$_2$ interface[40-42], $W_{eff} \approx W + 2t_{ox}$ is the effective width of the heated region at the SiO$_2$/Si interface, and $k_{Si} \sim 100$ Wm$^{-1}$K$^{-1}$ is the thermal conductivity of the doped Si substrate. We note eq. 3 includes fringing heat loss from the narrow GNRs, is mathematically similar to the capacitance expression in eq. 1, and was verified against finite-element simulations in Ref. 2.

The solid line in Figure 3c represents the model above with $W/L \sim 35/300$ nm, $t_g \sim 1.25$ graphene layers (averages for our samples, i.e. $\sim$0.42 nm), $t_{ox} = 90$ nm, and $R_{Cox} \sim 10^{-8}$ m$^2$KW$^{-1}$. Additional parameters of our model are the thickness of the metal electrode ($t_m$) and the thermal conductivities of the graphene, oxide, metal, and silicon substrate ($k_g$, $k_{ox}$, $k_m$, $k_{si}$), as described in Section F of the Supplement. The error bars and dashed lines estimate the effect of uncertainties on our extraction and model calculations, respectively. For instance, the variability of the plotted data is partly attributed to uncertainty in device dimensions, GNR-substrate thermal coupling, thermal conductivity of the devices, contact resistance, and breakdown temperature. (More discussion is also given in the Supplement.) With the parameters above, the percentage contribution of the first, second and third terms in eq. 3 are 65% (thermal resistance of SiO$_2$ including fringing heat loss), 34% (thermal resistance of graphene-SiO$_2$ interface), and 1% (thermal resistance of silicon substrate). Although the last term can usually be ignored[2], we include here all terms to highlight that their individual contributions depend strongly on device dimensions and oxide thickness. For carbon nanotubes[39] or extremely narrow GNRs, the graphene-SiO$_2$ thermal interface will dominate (also see Figure S5b).

We note that the current densities obtained for GNR interconnects in this work are higher (for a given resistivity and width) than those previously achieved, and reach ~$2 \times 10^9$ A/cm$^2$, as shown in Figure 3c. We attribute this to two advances in our understanding and thermal engineering of such small nanostructures. First, the GNRs here are shorter than previous devices[14, 38], being only slightly longer than the thermal healing length ($L_H$ ~ 0.1-0.2 μm) and enabling partial cooling through the metal contacts. Second, the GNRs in this work have been deliberately placed on a thinner oxide (~90 nm) vs. the ~300 nm used in previous studies[2, 14, 38]. The thinner oxide reduces the thermal resistance of these devices (see eq. 3 and Figure S5) for a given GNR width, and is an important factor enabling the higher current densities reached. A similar effect could be achieved by placing GNRs on other thin films with higher thermal conductivity ($k_{ox}$), or lower GNR-substrate thermal interface resistance ($R_{Cox}$). Such suggestions are consistent with recent measurements of larger graphene devices (not GNRs) on nanocrystalline diamond[43] (higher $k_{ox}$), and on BN substrates (possibly lower $R_{Cox}$ due to smoother graphene-BN interface)[44].

Before concluding, we also present measurements of our GNRs from CVD-grown graphene performed at the opposite end of the temperature spectrum, down to ~1.7 K, as shown in Figure 4 and Supplementary Figure S6. In general, we observed quantum-dot (QD), Fabry-Perot (FP) or universal conductance fluctuation (UCF) behavior, depending on the contact resistance ($R_C$) and quality of the devices (e.g. edges, impurities). Figure 4 displays such measurements for a short GNR ($L$ ~ 52 nm, $W$ ~ 35 nm), indicating QD-like behavior akin to previous observations in single- and bi-layer GNRs patterned on exfoliated graphene[31, 45-47].

The black curve in Figure 4b shows the modulation of zero-drain-bias conductance ($G$) as a function of back-gate $V_G$. The measurement is performed with a conventional lock-in technique with an excitation voltage $V_{ac}$ = 360 μV, corresponding to an electron temperature $T_e$ ~ 4.2 K (cryostat temperature 1.7 K). An electron overcomes a charging energy to be added into the QD, estimated as $E_C = e\Delta V_D$ ~ 2 meV ($\gg k_B T_e$), where $e$ is the elementary charge (Figure 4e). The Coulomb peaks in Figure 4b are a function of the electron distribution and can be described[47, 48] by $G(V_G) \sim \cosh^{-2}(\eta e(V_G - V_{G,P})/2.5 k_B T_e)$ for $\Delta E < k_B T_e < E_C$. Here, $\eta = \Delta V_D / \Delta V_G$ is the "gate factor", $V_{G,P}$ is a Coulomb peak gate voltage, $\Delta E$ is the spacing between neighboring single-particle levels in a QD, and $\Delta V_G$ is the gate voltage difference between adjacent Coulomb peaks. Gray oscillations in Figure 4b are fitted to the measured black curve using the above equation ($T_e = 4.2$ K), confirming that the conductance variations originate from Coulomb oscillations in a QD.

We can also estimate the size of the principal QD in Figures 4d and 4e. The width of the *N*-th diamond is $\Delta V_{G,N} = e/C_G$ for single electron tunneling, where $C_G$ is the gate capacitance. We obtain $\Delta V_G = 73 \pm 13$ mV for the Coulomb peaks in Figure 4b, resulting in $C_G = 2.2 \pm 0.4$ aF. On the other hand, $C_G$ can be also estimated from the device geometry including the effect of fringing fields at the GNR edges, $C_{ox} = 1.4$ mF/m$^2$ per unit area from eq. (1) with $W = 35$ nm. This allows us to estimate[47] the size of the QD as $C_G/C_{ox} = 1.6 \pm 0.27 \times 10^3$ nm$^2$, which yields $L_{QD} = 49 \pm 9$ nm (Figure 4d). The estimated length of the QD is near to the physical length of the GNR, indicating that one QD spans most of this particular GNR. The existence of an additional superimposed oscillation with a much larger period in Figure 4a-b could be attributed to a secondary coupled QD approximately ~10 times smaller in size[45, 46].

The observation of QD behavior spanning most of the GNR indicates that some of the shorter GNRs from CVD graphene are relatively defect-free quantum systems, although they do remain limited by their contacts. The presence of defects, grain boundaries and edge roughness[22, 31, 45] in longer ribbons, however, can distort the transport along the channel. Among our longer GNRs ($L > 100$ nm), some have also demonstrated F-P-like or UCF conductance oscillations (Figure S6 in Supplement) and others show multiple QDs in series.

In summary, we examined the fabrication, electrical and thermal behavior of GNR interconnects from CVD-grown graphene, a fundamental step towards their integration into large-scale applications. The GNRs presented here have low-field mobility and Raman signatures comparable to GNRs obtained by other methods. At high-field, small adjustments in thermal engineering such devices allow us to reach some of the highest current densities reported for any graphene interconnects ($>10^9$ A/cm$^2$). At low-temperatures, these GNRs display QD- or UCF-like transport behavior, depending on their dimensions and conductance levels. Transport in relatively short GNRs ($L < 100$ nm) appears dominated by contacts rather than by edge roughness, defects or grain boundaries. This work presents a unified view of low-field to high-field transport in GNRs over a very wide temperature range, and serves to identify remaining challenges which include reducing variability, surface impurities and contact resistance.

## ■ ASSOCIATED CONTENT

**Supporting Information.** Details of fabrication process and microscopy of GNRs; additional current-voltage characteristics, effect of GNR width on fringing capacitance, complete details of electrical and thermal models; additional low-temperature data and analysis. This material is available free of charge via the Internet at http://pubs.acs.org

## ■ ACKNOWLEDGMENT

This work was in part supported by the ARO PECASE Award, AFOSR and ONR Young Investigator Program (YIP), and the NSF. C.M.N. acknowledges support by the NSF-REU program. The authors also acknowledge useful discussions with Prof. Nadya Mason.

## ■ REFERENCES

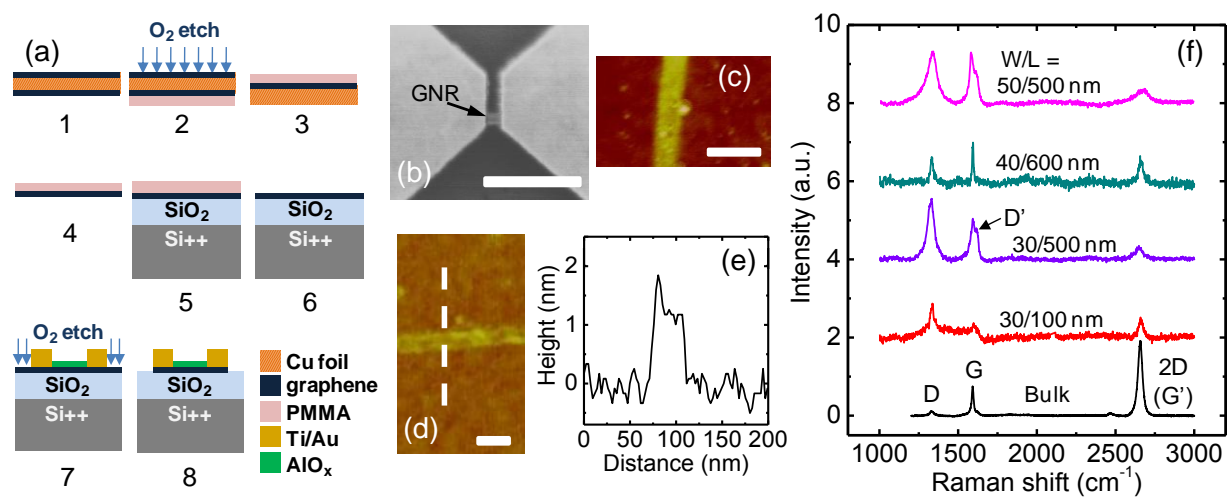

**Figure 1. (a)** Schematic of graphene growth and GNR fabrication process. **(b)** Scanning electron microscope (SEM) image of a GNR ($W \sim 75$ nm, $L \sim 110$ nm) between two Ti/Au electrodes; scale bar = 1 μm. **(c)** Atomic force microscope (AFM) image of AlO$_x$ strip covering a GNR ($W \sim 60$ nm); scale bar = 100 nm. Also see Figure S1. **(d)** AFM image of a GNR ($W \sim 35$ nm) after removal of the top AlO$_x$ strip; scale bar = 50 nm. **(e)** Cross section of AFM profile along dashed line in (d). Apparent topographic height in the 1-1.2 nm range (in air, including possible residue from fabrication) suggests the GNR is most likely monolayer, although bilayer cannot be ruled out[1, 49]. **(f)** Raman spectra (633 nm) for bulk CVD graphene (bottom curve) and several GNRs, spaced for clarity. The initial CVD graphene is predominantly monolayer (narrow 2D peak width $\sim 35$ cm$^{-1}$), while the D and D' bands of GNRs are more prominent due to the presence of edges.

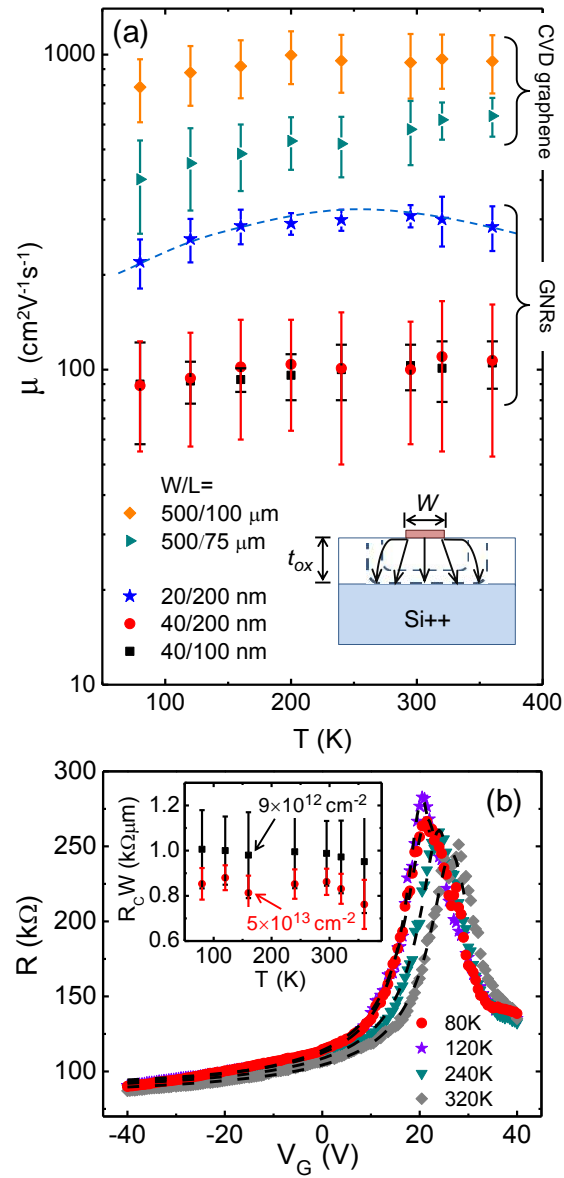

**Figure 2. (a)** Low-field hole mobility vs. temperature for GNRs from CVD graphene (three lower data sets) and large-area CVD graphene devices (two upper sets). All data shown at a charge density $p = 5 \times 10^{12}$ cm⁻². The fit[26] to experimental data takes into account thermally generated carriers ($n_{th}$), puddle charge ($n_{pd} = 1\sim4\times10^{12}$ cm⁻²) and contact resistance ($R_C W = 0.5\sim1$ kΩ·μm). For GNRs we include effect of fringing fields (see inset and text). Error bars show upper and lower bounds of $\mu$ extraction from the least-squares fit[26] with $R^2 \geq 0.9$ (see Supplement Section C). Dashed blue line is a guide for the eye. **(b)** Example of fitting GNR data (symbols) for $W \sim$ 20 nm device with the model (lines) at $V_{DS} = 50$ mV. The inset shows the contact resistance fit for the same temperatures, at two carrier densities.

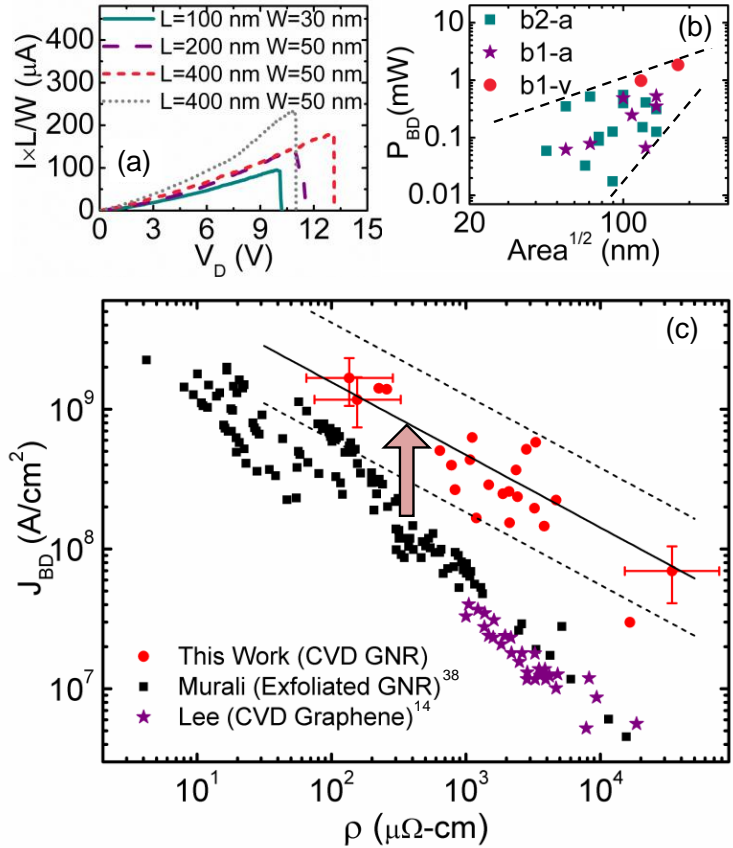

**Figure 3.** High-field properties of GNRs from CVD graphene. **(a)** Current-voltage measured up to electrical breakdown. The first three devices in the legend are GNRs capped by AlO$_x$, the last is uncapped. **(b)** Maximum power at breakdown increases approximately as square root of device area (see text). **(c)** Maximum current density vs. resistivity at breakdown for CVD GNRs on 90 nm SiO$_2$ (this work), GNRs patterned from exfoliated graphene (1 to 5 layers) on 300 nm SiO$_2$ (Murali et al.[38]), and large-area CVD graphene interconnects (10 to 20 nm thickness) on 300 nm SiO$_2$ (Lee et al.[14]). Our devices reach higher $J_{BD}$ in part due to better heat dissipation on the thinner (90 nm) oxide and smaller dimensions of the GNRs. The block arrow symbolizes this size effect. Representative error bars account for the uncertainty in $R_C$, $W$ and thickness $t_g$ of our GNRs. Lower dashed line represents the model (see text) assuming GNRs are 3-layers thick, have aspect ratio $W/L$ = 60/500 nm, graphene-oxide interface thermal resistance $R_{Cox}$ ~ 5×10$^{-8}$ m$^2$KW$^{-1}$ and thermal conductivity $k_g$ = 50 Wm$^{-1}$K$^{-1}$. Upper dashed line assumes monolayer graphene, $W/L$ = 15/100 nm, $R_{Cox}$ ~ 5×10$^{-9}$ m$^2$KW$^{-1}$ and $k_g$ = 500 Wm$^{-1}$K$^{-1}$. The solid line is obtained with $W/L$ = 35/300 nm, $R_{Cox}$ ~ 10$^{-8}$ m$^2$KW$^{-1}$ and $k_g$ ~ 100 Wm$^{-1}$K$^{-1}$, the latter being consistent with previous work on GNRs from unzipped nanotubes[2].

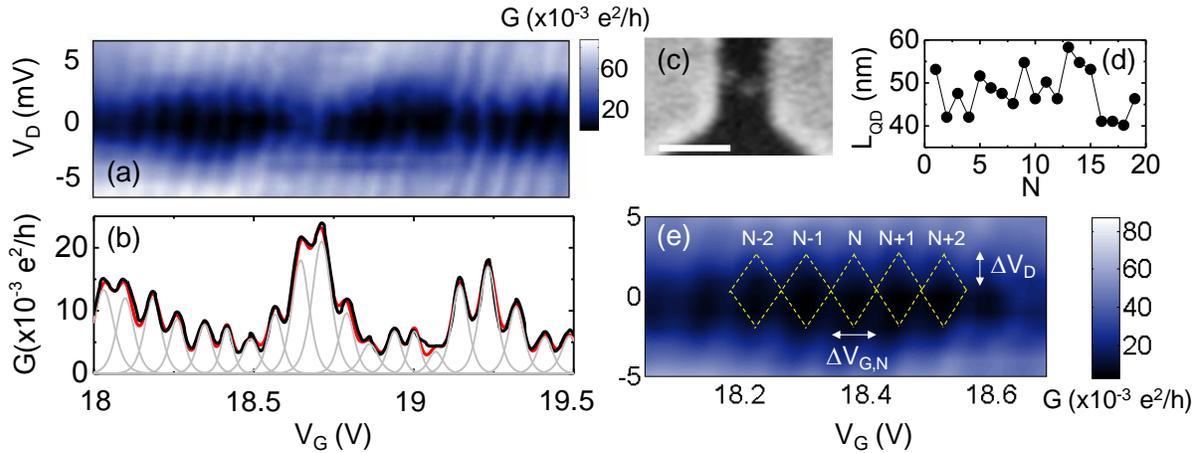

**Figure 4.** Low-temperature ($T$ = 1.7 K) measurement of a GNR ($L/W$ ~ 52/35 nm) from CVD graphene showing quantum dot (QD) behavior limited by the GNR-metal contacts. **(a)** Conductance map as a function of $V_G$ and $V_D$. **(b)** Conductance profile (black curve) at $V_D$ = 0 V. Gray curves are model including thermal broadening (see text). Red curve is the sum of all gray curves, in good agreement with the measured results. **(c)** SEM image of the GNR; scale bar = 100 nm. **(d)** Length of QD estimated from each peak $N$ is similar to the physical length of the GNR, indicating one QD spans most of the GNR channel. **(e)** Zoomed conductance map, where $\Delta V_{G,N}$ is the width of the $N$-th Coulomb diamond. $\Delta V_D$ is the drain voltage corresponding to the charging energy of a single electron (see text).